\documentclass[twocolumn,showpacs,preprintnumbers,amsmath,amssymb,prb]{revtex4-1}
\usepackage{upgreek}
\usepackage[latin1]{inputenc}
\usepackage{graphicx}
\usepackage{dcolumn}
\usepackage{bm}
\usepackage{color}
\usepackage{braket}

\begin{document}

\preprint{}

\title{Exciton spin dynamics and photoluminescence polarization of CdSe/CdS
dot-in-rod nanocrystals in high magnetic fields}

\author{B. Siebers$^{1}$, L. Biadala $^{1}$, D.R. Yakovlev$^{1,2}$, A.V. Rodina$^{2}$,
T. Aubert $^{3,4}$, Z. Hens  $^{3,4}$,    and M. Bayer$^{1,2}$}

\affiliation{$^{1}$Experimentelle Physik 2, Technische Universit\"at Dortmund, 44221 Dortmund,
Germany} \affiliation{$^{2}$Ioffe Physical-Technical Institute, Russian Academy of Sciences, 194021
St. Petersburg, Russia} \affiliation{$^{3}$Physics and Chemistry of Nanostructures, Ghent University,
9000 Ghent, Belgium} \affiliation{$^{4}$Center for Nano- and Bio-photonics, Ghent University, 9000
Ghent, Belgium}

\date{\today}

\begin{abstract}
The exciton spin dynamics and polarization properties of the related emission are investigated in colloidal CdSe/CdS dot-in-rod (DiR) and spherical core/shell nanocrystal (NC) ensembles by magneto-optical photoluminescence (PL) spectroscopy in magnetic fields up to 15~T. It is shown that the degree of circular polarization (DCP) of the exciton emission induced by the magnetic field is affected by the NC geometry as well as the exciton fine structure and can provide information on nanorod orientation. A theory to describe the circular and linear polarization
properties of the NC emission in magnetic field is developed. It takes into account phonon mediated coupling between the exciton fine structure states as well as the dielectric enhancement effect
resulting from the anisotropic shell of DiR NCs. This theoretical approach is used to model the experimental results and allows us to explain most of the measured features. The spin dynamics of the dark excitons is investigated in magnetic fields by time-resolved photoluminescence. The results highlight the importance of confined acoustic phonons in the spin relaxation of dark excitons. The bare core surface as well as the core/shell interface give rise to an efficient spin relaxation channel, while the surface of core/shell NCs seems to play only a minor role.
\end{abstract}

\maketitle

\section{Introduction}

The on-going research and interest in colloidal NCs over the past three decades is partly driven by the prospect of building low cost, high quality and highly efficient lighting or light harvesting
devices,\cite{Sanderson2009,Wood2010,CoeSullivan2009,Talapin2010a,Shirasaki2013} such as colloidal NC based light emitting diodes \cite{Rogach2008,Caruge2008} or solar cells,\cite{Luther2008,Sargent2012}
and partly by the opportunities opened up due to the high expertise in synthesis methods, enabling one to produce NCs of almost arbitrary size and shape. A vast improvement regarding the optical properties of NCs has been the introduction of a shell, resulting in effective passivation of
surface trap states of the NC core, thereby considerably improving the quantum yield.\cite{Reiss2009}
It has been also shown that the shell provides the possibility to manipulate the confinement of electrons and holes separately enabling to tailor their wavefunction overlap and thereby to
control their interaction.\cite{Kim2003,Brovelli2011,Raino2011,Javaux2013} In that way further improvement of the optical properties of NCs can be obtained.

At low temperatures the light emission from NCs is predominantly determined by the exciton fine structure, which in wurtzite type NCs arises from the interplay of the crystal lattice anisotropy, the
NC shape and the electron-hole exchange interaction.\cite{Efros1996} Due to its importance for the radiative emission of NCs the exciton fine structure has been studied theoretically and
experimentally for various NC materials, sizes and shapes, such as spherical core-only and core/shell NCs,\cite{Furis2006,Biadala2010a,Labeau2003,DeMelloDonega2006} nanorods
\cite{Shabaev2004,LeThomas2005,Zhao2007} and dot-in-rods (DiRs).\cite{Aguila2014,Biadala2014}

In spherical NCs the most crucial feature of the exciton fine structure is that the ground exciton state has total angular momentum $F=\pm 2$ and is therefore optically forbidden in the electric-dipole approximation, i.e. is a dark state. Optically active states with $F=\pm 1$ are split from the dark state typically by a couple of meV to higher energies. This splitting is often referred to as the bright-dark energy splitting and its magnitude depends on the electron-hole exchange
interaction.\cite{Efros1996} Notably it has been shown, that by increasing the CdS shell thickness due to the electron leakage into the shell the electron-hole exchange interaction and thereby the magnitude of the bright-dark splitting can be controlled.\cite{Brovelli2011} In contrast, recent experiments on DiRs indicate, that the shape of the CdS shell does not significantly affect the level order in the exciton fine structure for the core sizes of interest here.\cite{Aguila2014,Biadala2014} While in bare anisotropic structures, such as nanorods, a change in the order of the valence subbands may occur at a certain diameter, leading to an exciton ground state with total angular momentum of $F=0$,\cite{Shabaev2004,LeThomas2005,Louyer2011} the fine structure of DiRs appears to be similar to the spherical case. The order of the fine structure states is crucial for the photoluminescence (PL) polarization of the NC, especially at low temperatures, where the thermal energy is smaller than the fine structure splitting and mostly the lowest lying exciton states contribute to radiative recombination. While for DiR NCs high degrees of linear polarization of the PL have been experimentally observed at elevated
temperatures,\cite{Talapin2003,Pisanello2010,Hadar2013} corresponding studies at low temperatures are still missing.

The recombination dynamics of NCs is controlled by transitions between exciton fine structure states involving energy and spin relaxation. In order to conserve energy and angular momentum, these transitions are accompanied by absorption or emission of phonons. However, since phonons alone do not flip spins additional mechanisms for spin relaxation are required. Understanding these mechanisms is important because it enables design of spin-preserving environments for future spin-based devices in information storage.\cite{Kroutvar2004,Chappert2007} Colloidal NCs, due to their high versatility regarding synthesis and processing as well as their unique and widely tunable optical properties are appealing candidates for such applications.\cite{Bussian2009,Talapin2010a} Thus,
comprehensive knowledge about the linkage between NC size, shape and material properties and exciton spin dynamics is highly desirable. In recent experiments, exciton spin relaxation and its dependence
on the NC shape have been addressed and possible mechanisms involved have been discussed.
\cite{Scholes2006,Scholes2006a,Kim2006,He2008,Kim2009,Wong2009} However, in these studies the authors consider only exciton relaxation for bright-to-bright and bright-to-dark states and the experiments have been conducted at elevated temperatures. Transitions between the dark exciton ground states, which become important at low temperatures and occur in the presence of external perturbations, such as an external magnetic
field,  have not been discussed so far. As pointed out in Ref.~\onlinecite{Poem2010}, the dark
exciton due to its long lifetime is a promising candidate for long-lived spin storage applications.

In this paper, we present a detailed study of exciton spin dynamics and PL polarization in CdSe/CdS core/shell NC ensembles subject to an external magnetic field. We report on how the spin dynamics between
the dark exciton ground states split by the magnetic field is affected by the size and shape of the CdS shell. To this end we measure the time-resolved degree of circular polarization (DCP) as function of temperature and magnetic field for NCs covering a wide range of shapes and
sizes, including DiRs as well as spherical core-only and core/shell NCs. Interestingly we find that the core/shell interface strongly affects the spin relaxation rate, whereas the shell shape and size
have only slight impact in this regard. These results highlight the crucial role of the core/shell interface for the optical properties of NCs, especially for the relaxation and recombination dynamics of dark excitons. Moreover, our results suggest that the spin flip is accompanied by the emission of confined acoustic phonons and remarkably NCs of spherical shape show a behavior similar to DiRs in this regard.

We also study the impact of the shell shape on the PL circular polarization induced by the magnetic field. On the one hand we show that by measuring the PL-DCP one can probe the orientation of DiR
ensembles, which after drop-casting on a substrate can show macroscopic alignment features in so-called ''coffee stain rings''.\cite{Nobile2009,Zavelani-Rossi2010} On the other hand, we analyze the dependence of the PL-DCP on magnetic field for DiRs and spherical core/shell NCs. Our results indicate, that the saturation level of the DCP in DiRs depends on the aspect ratio of the DiR
shell. We elaborate a theoretical approach for analyzing and predicting the polarization properties of NCs in the presence of an external magnetic field, which includes a dielectric enhancement effect
resulting from the anisotropic shell as well as phonon-mediated coupling of the dark exciton ground state to bright exciton states with radiative transitions parallel and perpendicular to the hexagonal
axis. Interestingly our model suggests that, although at high temperatures the PL of DiR may be polarized along the hexagonal axis, at low temperatures and in the presence of magnetic fields it can be polarized perpendicular to it.

The paper is organized as follows: In section \ref{experimentals} information on the synthesis of the NCs is given and the experimental setup is described. In section \ref{experimentalresults} the
experimental results are presented and discussed. Here, we start with a general discussion about how the orientation of wurtzite type NCs affects the circular polarization in magnetic field and we present
results of measurements, in which such orientation effects are observed. Subsequently we explain the experimental procedure for investigating the exciton spin dynamics and the equilibrium PL-DCP. Thereafter results on the spin dynamics and DCP are presented and discussed. In section
\ref{theorypart} the theory for describing the PL circular and linear polarization of DiRs in external magnetic field is presented and compared to the experimental data. The paper concludes with a summary and an outlook in section \ref{outlook}.

\section{Experimentals}
\label{experimentals}

The wurtzite CdSe cores and the CdSe/CdS DiR colloidal NCs were prepared following the protocol described in Ref.~\onlinecite{Carbone2007}. In this method, the use of phosphonic acids for the seeded growth of the CdS shell results in an anisotropic growth of the shell around the spherical
core. The spherical CdSe/CdS NCs were prepared starting from the same wurtzite CdSe cores by using the recently reported ''flash'' synthesis.\cite{Cirillo2014} This method makes use of oleic acid
instead of phosphonic acids, which promotes an isotropic growth of the CdS shell and results in a spherical morphology of the core/shell NCs. Parameters of the studied samples are collected in the Tables~\ref{table1} and \ref{table2} for the DiRs and spherical NCs, respectively.

\begin{table*}[t!]
\caption{Parameters of the studied CdSe/CdS DiR samples. $E_m$ is the energy of the PL maximum, $P_c^\text{sat}$ - saturation level of the DCP, estimated by fitting Eq.~\eqref{bdependence_eq} to the experimental data. $g_{ex}$ - exciton $g$ factor, $R_e$ - enhancement factor, and  $r=R_e\Gamma_{20}/\Gamma_{21}^0$ - parameter defined in Eq.~\eqref{bdependence_eq}.   Numbers in brackets are standard deviations.}
  \label{table1}
  \begin{tabular}{|c|c|c|c|c|c|c|c|c|c|}
  \hline
\#& Core diameter   & Shell length  & Shell width & Aspect ratio & $E_m$ (eV) & $P_c^\text{sat}$ & $g_{ex}$&$R_e$&$r$ \\
& $D$ (nm) & $L$ (nm) & $W$ (nm) & $L/W$ & & & & & \\
\hline
1 & $2.5$& $28.9\, (1.5)$ & $5.3\, (0.5)$ & 5.5 & 2.18 & -0.65&2.7 & 3.2&0.03\\
\hline
2 & $2.5$& $22\, (1)$ & $3.8\, (0.3)$&5.8&2.23&-0.55&2.7&3.3&0.13\\
\hline
3 & $2.5$ & $22.8\, (1.8)$ & $3.4\, (0.2)$&6.7&2.26&-0.49&2.7&3.4&0.19\\
\hline
\hline
4 & $3.2$& $60\, (5)$ & $4.1\, (0.3)$&14.6 &2.11& -0.63& 1.4&3.8&0.05\\
\hline
5 & $3.2$ & $32\, (1)$ & $4.1\, (0.3)$&7.3&2.12& -0.60&2.0&3.5&0.08\\
\hline
\hline
6 & $3.7$ & $20.4\, (1.3)$&$6\, (0.6)$&3.4&2.03&-0.57&1.4&2.8&0.16\\
\hline
7 & $3.7$ &$26.5\, (2.7)$ &$5\, (0.3)$&5.3&2.04&-0.63&1.5&3.2&0.06\\
\hline
\end{tabular}
\end{table*}

\begin{table*}
  \caption{Parameters of the studied spherical samples.}
  \label{table2}
  \begin{tabular}{|c|c|c|c|c|c|c|c|}
  \hline
\#& Core Diameter $D$ (nm) & Shell thickness $S$ (nm) & $E_m$ (eV) & $P_c^\text{sat}$ & $g_{ex}$&$R_e$ &r\\
\hline
R1 & $3.1$&  $2$ &  2.06 & -0.74&2.7&1 &0.02\\
\hline
\hline
R2 & $2.7$& $0$&2.32&-0.67&1.2&1&0.25\\
\hline
R3 & $2.7$  & $0.8$&2.18&-0.74&2.4&1&0.02\\
\hline
R4 & $2.7$ & $3.7$ &2.07& -0.67&2.8&1&0.25\\
\hline
R5 & $2.7$ & $5.3$&2.01& -0.60&0.9&1&0.5\\
\hline
\end{tabular}
\end{table*}

\begin{figure}[t!]
\centering
\includegraphics[width=\linewidth]{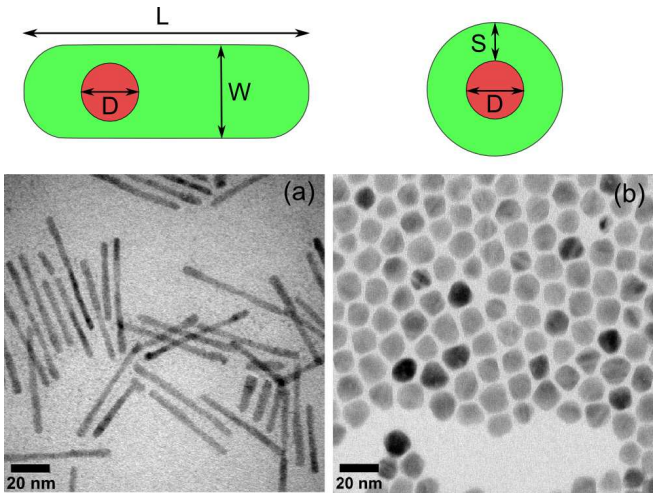}
\caption{(Color online) (a) Transmission electron microscope image of sample \#4. (b) Transmission electron microscope image of sample \#R5. The sketches on top illustrate the geometry of a DiR and a spherical core/shell NC.}
\label{figure1}
\end{figure}

Figure~\ref{figure1} shows typical transmission electron microscope (TEM) images of CdSe/CdS DiRs and spherical core/shell NCs. The geometrical parameters are illustrated by the top sketches. The sizes of the CdSe/CdS DiRs and spherical core/shell NCs were determined from such TEM images. The size of the bare CdSe cores was determined from the energy of the first peak in the absorption spectrum, using published data for the size
dependence of the optical transitions.\cite{Jasieniak2009}

For magneto-optical photoluminescence measurements, a solution of NCs in toluene with a concentration of a few $\mu$M was deposited by drop-casting onto a glass slide and afterwards dried. For low
temperature experiments the samples were mounted into an optical cryostat with direct optical assess to the sample via windows. The temperature could be varied between $T=2.2$~K and 20~K. The cryostat is equipped with a superconducting solenoid for
generating magnetic fields up to 15~T.  The magnetic field, $B$, was applied in Faraday geometry, i.e., parallel to the light wave vector and perpendicular to the sample substrate. The sample
was mounted on a holder top attached to a nanopositioner system, allowing for precise positioning in the three
spatial directions. NC photoluminescence was excited with a linearly-polarized pulsed diode laser (pulse
duration of about 100~ps), operating at a wavelength of 405~nm (photon energy of 3.06~eV) at a repetition rate of 1~MHz.
The photoluminescence was detected either with an avalanche photodiode
(time resolution of 100~ps) or spectrally-resolved by a monochromator connected to a nitrogen-cooled charge-coupled device camera. The overall time resolution of the setup was determined to be 250~ps. The circular polarization of the PL was analyzed using a quarter wave plate and a
Glan-Thomson-prism.

\section{Experimental results}
\label{experimentalresults}

\subsection{Effect of NC orientation on DCP}

\begin{figure}[t]
\centering
\includegraphics[width=\linewidth]{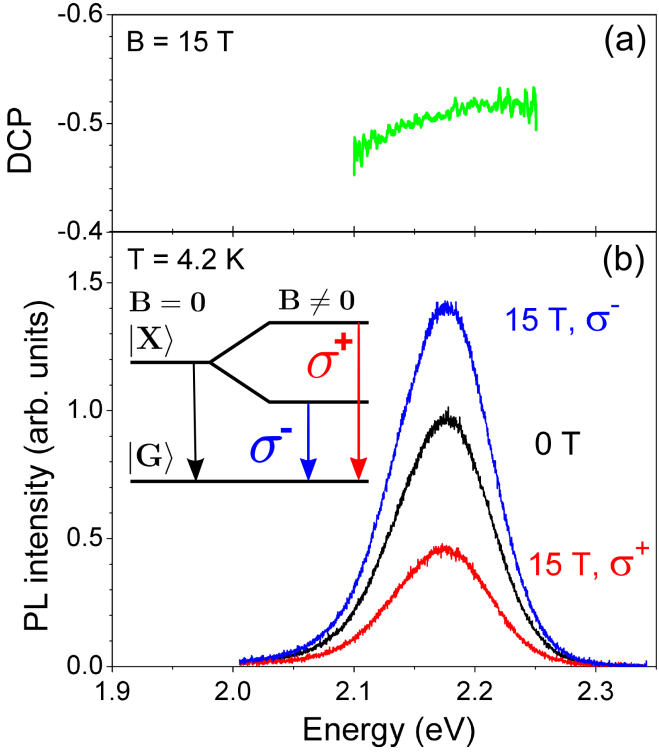}
\caption{(Color online) (b) Photoluminescence spectra of DiR sample \#1 at $B=0$~T and $15$~T. Red and blue curves are the $\sigma^+$  and $\sigma^-$  polarized components, respectively. The black curve corresponds to the unpolarized transition, as illustrated in the inset. The upper panel (a) shows the spectral dependence of the DCP. The insert shows the optical transitions between an exciton state $|\text{X}\rangle$ that is split in magnetic field into two spin sublevels and the unexcited ground state of the crystal $|\text{G}\rangle$.}
\label{figure2}
\end{figure}

Figure~\ref{figure2}(b) shows polarization-resolved PL spectra of sample~\#1 at $B=0$~T and 15~T. Due to the size distribution in the NC ensemble the PL linewidth is broadened up to approximately 80~meV
(i.e.~20 nm). While at zero magnetic field the PL is unpolarized, it exhibits a strong DCP of $-0.5$ (i.e. $-50\%$) at 15~T.  The degree of circular polarization $P_c$ is calculated as
\begin{equation}
P_c=\frac{I_{\sigma+}-I_{\sigma-}}{I_{\sigma+}+I_{\sigma-}},
\label{dcpintensities}
\end{equation}
where $I_{\sigma+}$ and $I_{\sigma-}$ are the PL intensities of the right- and left- circular polarized components, respectively. The spectral dependence of the DCP is shown in Fig.~\ref{figure2}(a). It changes only slightly across the PL band, which may be due to variation of the NC parameters or contribution of spectral diffusion. Here we will not go into the details of these variations so that we can use spectrally-integrated PL intensities for calculation of the DCP. The observed polarization is result of an exciton distribution among the Zeeman-split sublevels, as illustrated in the inset of Fig.~\ref{figure2}(b). In NCs with wurtzite crystal structure, due to the internal crystal anisotropy
the exciton spin is constrained to orientations along the hexagonal c-axis of the crystal.\cite{Johnston-Halperin2001} As a result, the magnitude of the exciton Zeeman splitting depends on the angle $\Theta$ between the crystal c-axis and the direction of the magnetic field:
$\Delta E = g_{ex}\mu_B\mathbf{B}\cdot\mathbf{c}=g_{ex}\mu_BB\cos\Theta$, where $g_{ex}$ is the exciton $g$ factor and $\mu_B$ is the Bohr magneton.  The intensities $I_{\sigma\pm}$ in Eq.~\eqref{dcpintensities} therefore also depend on the angle $\Theta$. The probability for detecting circular polarized light at an angle $\Theta$ with respect to the magnetic field axis is proportional to $(1\pm\cos{\Theta})^2$.\cite{Johnston-Halperin2001} Thus, for
an ensemble of NCs the $P_c(B)$ is derived by integrating Eq.~(\ref{dcpintensities}) over all possible angles
\begin{equation}
P_c(B)=\frac{\int_0^1 [I_{\sigma+}(x)-I_{\sigma-}(x)]f_\text{or}(x)dx}{\int_0^1 [I_{\sigma+}(x)+
I_{\sigma-}(x)]f_\text{or}(x)dx}.
\label{bdependence_gen}
\end{equation}
Here, $x=\cos{\Theta}$ and $f_\text{or}(x)$ is a weighting function
describing the orientation distribution in the ensemble.

Assuming a thermal distribution between the Zeeman-split sublevels in equilibrium, the calculated DCP is
\begin{equation}
P^{\text{eq}}_c(B) =\frac{\int_0^1 2x\tanh(\Delta E
/2k_BT)f_\text{or}(x) dx}{\int_0^1(1+x^2)f_\text{or}(x) dx}
\label{bdependence},
\end{equation}
where $k_B$ is the Boltzmann constant. For an ensemble of randomly oriented NCs the angle $\Theta$ between the crystal c-axis and the magnetic field direction is equally distributed over all angles
between $0^{\circ}$ and $90^{\circ}$, i.e. $f_\text{or}(x)=1$. In this case, the integration over all angles $\Theta$ leads to a theoretical limit for $P^{\text{eq}}_c(B)$ of 0.75 (i.e. $75$~\%). However, for NCs with an anisotropic shape, such as nanorods or DiRs the situation can be different. For DiR NCs it is known that the c-axis of the core is typically parallel to the long axis of the rod.\cite{Talapin2003} Several methods have turned out to be suitable to prepare nanorod ensembles
with macroscopic alignment.\cite{Jana2004,Wang2004,Carbone2007,Ghezelbash2006,Ryan2006}
Magnetooptical spectroscopy can be used for reliable characterization of the nanorod orientation in the ensemble by measuring $P^{\text{eq}}_c(B)$, as one can see from
Eq.~(\ref{bdependence}). In the case when all nanorods are oriented perpendicular to the magnetic field direction, i.e. $\Theta=90^{\circ}$, the expected DCP at any magnetic field $P^{\text{eq}}_c(B)=0$. On the contrary, when all nanorods are oriented along the magnetic field, i.e. $\Theta=0^{\circ}$, the resulting $P^{\text{eq}}_c(B)$ is maximal and depends only on the ratio $\Delta E/2k_BT$. This case corresponds to $f_\text{or}(x)=\delta(x-1)$. For fully spin polarized excitons $P^{\text{eq}}_c(B)=1$.

As a proof of principle we used this method to probe the nanorod orientation in ensembles prepared by the drop-casting method. When a solution of highly concentrated NCs is drop-casted on a substrate,
its distribution on the substrate is subject to the so called ``coffee stain effect''.\cite{Deegan1997,Nobile2009,Zavelani-Rossi2010} Due to this effect the nanoparticles accumulate in the outermost part of the drying droplet, leading to the formation of a coffee stain ring, where the NCs are densely packed. As it was shown in Ref.~\onlinecite{Nobile2009} in these coffee stain rings nanorods can align macroscopically. Namely, they tend to be oriented parallel to each other and parallel to the substrate, whereas in the regions outside the ring the nanorods show no specific order and their orientation is random.

\begin{figure*}[t]
\centering
\includegraphics[width=\linewidth]{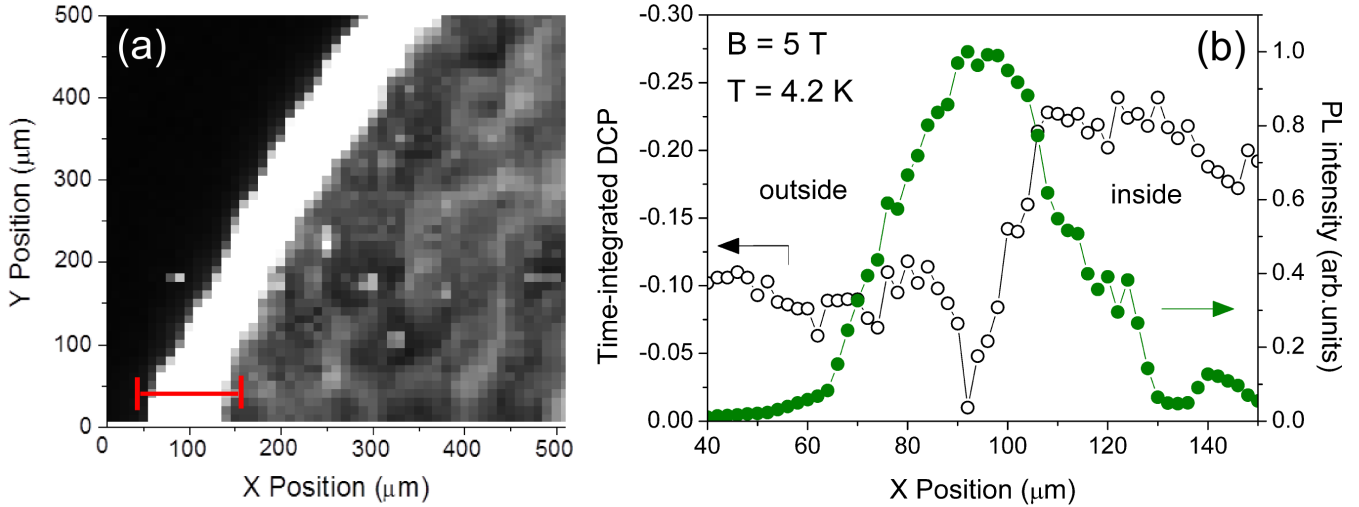}
\caption{(Color online) (a) Confocal laser scanning microscope image of a coffee stain ring section of sample \#5. The red line indicates the region, which was chosen for the linear scan shown in panel (b). Here, the spectrally-integrated PL intensity is depicted by the solid green circles and the time-integrated DCP by the open circles.}
\label{figure3}
\end{figure*}

Figure~\ref{figure3}(a) shows a PL intensity scan of a coffee stain ring section for sample~\#5 measured by a confocal laser scanning microscopy with a spatial resolution of $10~\mu$m. The spectrally-integrated PL intensity is plotted using a logarithmic grey scale with the brightness increasing with intensity. The high intensity within the ring supports the dense packing of nanorods in this area. In the inner part of the ring (grey area) the nanorod density is substantially smaller. Outside the ring the intensity is almost zero but still measurable (black area). To obtain more insight the PL intensity profile along the red line in Fig.~\ref{figure3}(a) is shown by the solid circles in
Fig.~\ref{figure3}(b). The time-integrated DCP at $B=5$~T for the scan across the ring is also shown in this figure by the open circles. It is remarkable, that the DCP value being $-0.23$ inside the ring drops almost to zero on the ring. This decrease clearly evidences the nanorod alignment parallel to the substrate surface resulting from the vanishing exciton Zeeman splitting for the magnetic field perpendicular to the c-axis of the majority of nanorods. Note that the time-integrated DCP in general is lower than $P_c^{eq}$. However, if the spin relaxation time is much shorter than the exciton recombination time this difference becomes
negligible.\cite{Johnston-Halperin2001, Liu2013} It will be shown below that this is always the case for the samples investigated here.

The technique presented is suitable to yield information about the nanorod orientation only with respect to the magnetic field axis perpendicular to the sample substrate.  Although it cannot compete with electron microscopy in terms of resolution, it has noteworthy advantages. While for TEM imaging special carbon grids, which are transparent
to the electron beams, have to be used, the method we propose is applicable to a wide range of substrate materials and can be used for device characterization. Furthermore, while the area available to TEM is typically limited to a few microns, the optical technique can be applied to much larger scales. Its spatial resolution is limited to several hundred nanometers due to light diffraction, depending on the laser wavelength and the numerical aperture of the optical excitation and detection system.

\subsection{Experimental investigation of equilibrium DCP and exciton spin dynamics}

In the preceding section we have highlighted the impact of the nanorod orientation on the magnetic-field-induced DCP in DiR nanocrystals. Therefore, when investigating the magneto-optical properties, this aspect has to be taken into account. In order to ensure comparability between
different samples, all measurements presented in the following were performed in the center of the dried droplets, where no specific order of the DiR particles is expected.

\begin{figure*}[t]
\centering
\includegraphics[width=\linewidth]{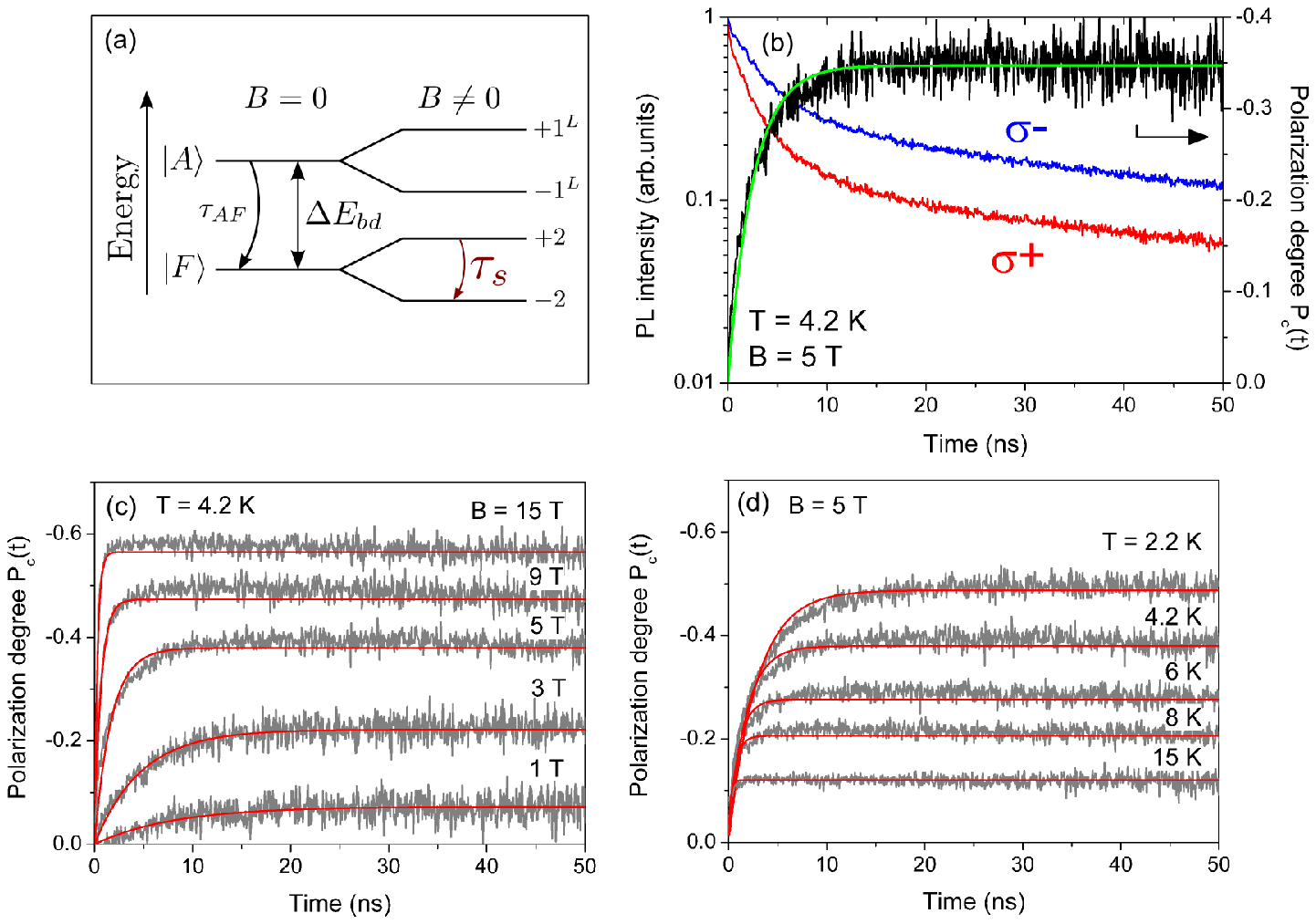}
\caption{(Color online) (a) Scheme of the lowest exciton fine structure states for zero and non-zero magnetic field. (b) Typical polarization-resolved PL decay and resulting time-resolved DCP for sample \#1. The green line is a fit according to Eq.~(\ref{fit}) with $\tau_s=2.5$~ns. (c) Magnetic field dependence and (d) temperature dependence of the time-resolved DCP. The red curves are fits according to Eq.~(\ref{fit}).}
\label{figure4}
\end{figure*}

The internal crystal structure and NC shape anisotropy, in combination with the electron-hole exchange interaction, lift the eight-fold degeneracy of the exciton ground state, leading to an exciton fine structure consisting of five energy levels, which are distinguished by their total angular momentum.\cite{Efros1996} At low temperatures it is typically sufficient to consider only the lowest exciton states when discussing the optical properties of NCs. It was shown in previous studies \cite{Aguila2014} that the exciton fine structure of DiRs for the core sizes of interest here is quite similar to the case of spherical core/shell NCs. A sketch of the lowest fine structure states is shown in
Fig.~\ref{figure4}(a) without and in presence of an external magnetic field. The lowest exciton ground state $\ket{F}$ has total angular momentum of $|F|=2$ and is optically forbidden. It lies below an optically allowed state $\ket{A}$ with $|F|=1$. The bright-dark energy splitting $\Delta E_{bd}$ is strongly dependent on the shell thickness, because a change in shell thickness leads to a modified electron-hole exchange interaction. This aspect was investigated in several experimental studies.\cite{DeMelloDonega2006,Brovelli2011,Biadala2014} For the samples we investigate here the magnitude of $\Delta E_{bd}$ lies in the range of $2-5$~meV.\cite{Biadala2014} At low temperatures PL emission occurs mainly from the dark state, due to efficient spin flip scattering from the bright to the dark exciton state, which occurs on picosecond timescale.\cite{Labeau2003,Hannah2011} Dark exciton radiative recombination is a long standing question and its origin has not been fully disclosed. Several reports evidenced a phonon-assisted mechanism involving longitudinal optical and longitudinal acoustic phonons. This, however, cannot explain some peculiar experimental observations, such as presence of the zero phonon line \cite{Nirmal1994,Efros1996,Biadala2009,Aguila2014} or shortening of the PL decay with increasing external magnetic field.\cite{Efros1996}

In the presence of a magnetic field the exciton states are split into Zeeman doublets with well-defined spin orientation. At low temperatures the establishment of thermal equilibrium population among the exciton spin sublevels involves the bright-dark spin relaxation characterized by a time $\tau_{AF}$ and the spin relaxation between the Zeeman sublevels of the bright and dark excitons. Since exciton relaxation between $\ket{+2}$ and $\ket{-2}$ requires a simultaneous spin flip of electron and hole, it is expected that this process is less efficient compared to the bright to dark exciton relaxation, i.e. $\tau_s\gg \tau_{AF}$, where $\tau_s$ is the spin relaxation time of the dark exciton.\cite{Biadala2010a} The spin relaxation between the $\ket{+2}$ and $\ket{-2}$ states is the main focus of this study.

The spin dynamics can be investigated by analyzing the time-resolved DCP
$P_c(t)$, as shown in Fig.~\ref{figure4}(b). After pulsed excitation, the $\sigma^+$ and $\sigma^-$ polarized components are traced as function of time. The time-resolved DCP, which is shown by the black curve, is calculated from the experimental data after Eq.~(\ref{dcpintensities}). Right after excitation the PL is unpolarized, because the photoexcited
excitons being excited with linearly polarized (unpolarized) laser light with large excess energy do not have a preferential spin orientation. Subsequently the excitons relax into the energetically lower lying spin states until they achieve thermal equilibrium, as evidenced by the plateau of the DCP in Fig.~\ref{figure4}(b). The characteristic time of the spin relaxation process involved can be quantified by determining the characteristic rise time of the time-resolved DCP.\cite{Liu2013} In the following we analyze the dependence of the PL polarization and spin dynamics on magnetic field, temperature and
shape of the NCs. To that end, we fit the time-resolved DCP curves with the equation
\begin{equation}
P_c(t)=P_c^{\text{eq}}[1-\exp(-t/\tau_s)],
\label{fit}
\end{equation}
where $P_c^{\text{eq}}$ denotes the DCP in thermal equilibrium and
$\tau_s$ the spin relaxation time of the dark exciton. A typical fit is shown by the green curve in Fig.~\ref{figure4}(b).

Figures~\ref{figure4}(c) and \ref{figure4}(d) show the $P_c(t)$ at various $B$ and $T$ for sample~\#1. On the one hand, as the temperature is kept constant and the magnetic field is increased, the DCP
saturates at higher values, due to the increased magnitude of the Zeeman splitting. On the other hand, the spin relaxation process becomes more efficient, as can be seen from the faster DCP increase at higher fields. If the magnetic field strength is fixed and the temperature is increased, the saturation level decreases due to thermal population of the higher lying Zeeman-sublevel resulting in a lower degree of spin polarization. In contrast the spin relaxation efficiency increases with increasing temperature, as one can see from the shortening of the DCP rise time in Fig.~\ref{figure4}(d). In the following we will first discuss the spin dynamics and will later focus on the dependences of $P_c^{\text{eq}}$ on magnetic field and NC shape.

\subsection{Exciton spin dynamics}

In this part the exciton spin dynamics and its dependence on magnetic field, temperature and NC shape is discussed. First, we want to stress that the observed dynamics of the $P_c(t)$ curve of sample \#1 in Fig.~\ref{figure4}(b) originates predominantly from relaxation between the dark exciton $F=\pm 2$ states. The plateau, which is reached at $t\approx 15$~ns clearly indicates thermally equilibrated excitons among the Zeeman-split fine structure states. The characteristic rise time obtained from the fit of the experimental $P_c(t)$ dependence with Eq.~\eqref{fit} is $\tau_s= 2.5$~ns in this case ($B=5$~T and $T=4.2$~K). The thermalization of excitons among the bright and dark states was studied in detail in Ref.~\onlinecite{Hannah2011}. Notably, this study revealed a strong size dependence of the thermalization time that was found to depend linearly on the square of the NC radius. At temperature $T=2.6$~K for NCs with a radius below $3$~nm the thermalization was found to occur on timescales $\tau_{AF}<100$~ps. Considering the range of sizes in our set of samples, we expect the bright to dark thermalization to occur on a similar timescale. This is more than one order of magnitude shorter than the typical spin relaxation times we determine in our experiments, especially at low magnetic fields. We believe that we are not able to resolve the dynamics on such a short timescale due to the limited resolution of our setup ($\approx 250$~ps). However, since $P_c(t)$ at $t=0$ is equal to zero, we suggest, that the relative population of the Zeeman-split bright states is equal, i.e. no thermalization takes place at such short timescales. As a result this equal population is transferred onto the dark states. Consequently, it is very likely that the dynamics of $P_c(t)$ at early times arises from the relaxation between the $\ket{+2}$ and $\ket{-2}$ dark exciton states.

\begin{figure}[t]
\centering
\includegraphics[width=\linewidth]{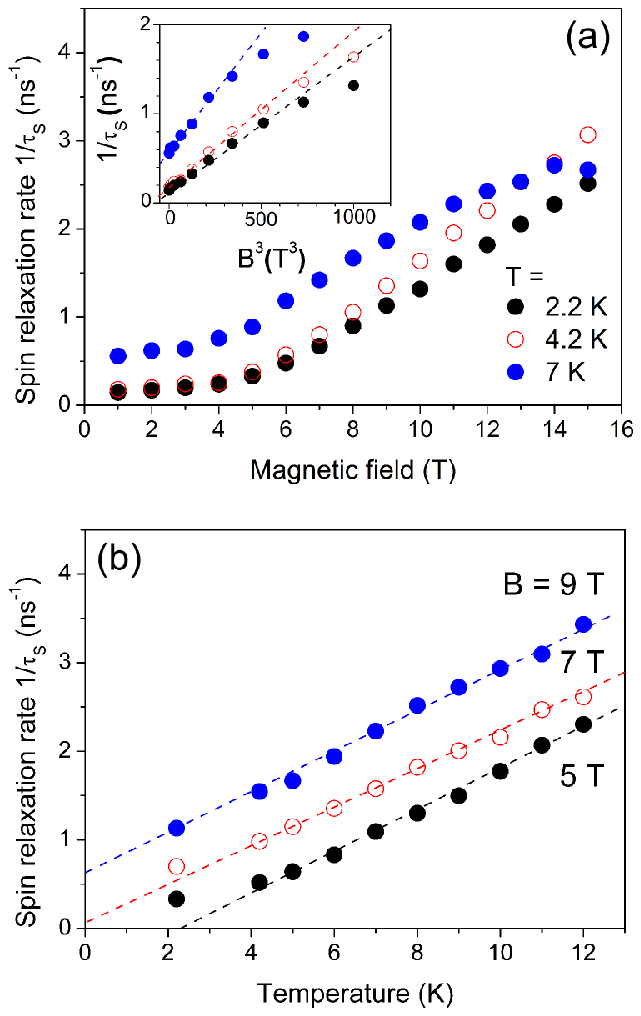}
\caption{(Color online) (a) Magnetic field dependence and (b) temperature dependence of the exciton spin relaxation rate $1/\tau_s$ for sample \#1. Dashed lines are guides to the eye.}
\label{figure5}
\end{figure}

Figures~\ref{figure5}(a) and \ref{figure5}(b) show the spin relaxation rate $1/\tau_s$, obtained for sample \#1 from the fit of the time-resolved DCP curves with Eq.~(\ref{fit}), as function of magnetic field and temperature. Examples of such fits are shown by the red lines in Figs.~\ref{figure4}(c) and \ref{figure4}(d). As can be seen in Fig.~\ref{figure5}~(a), the spin relaxation rate is strongly dependent on
the magnetic field. At $T=4.2$~K the rate increases from $1/\tau_s\approx 0.1$~ns$^{-1}$ to $3.3$~ns$^{-1}$ (this corresponds to a shortening of the relaxation time from $\tau_s\approx 10$~ns to $0.3$~ns) with increasing magnetic field from $B=0$~T to $15$~T. As one can see in the inset of Fig.~\ref{figure5}(a), at low magnetic fields the spin relaxation rate follows a cubic magnetic field dependence and it starts to deviate from this trend at $B\approx8$~T. At high fields the rate seems to depend linearly on magnetic field. The temperature dependence of the spin relaxation rate is a linear function, see
Fig.~\ref{figure5}(b). As the temperature is increased from
$T=2.2$~K up to $12$~K the spin relaxation rate increases by about one order of magnitude. The linear dependence on temperature points towards a phonon mediated process. 
For confined phonon energies much smaller than the thermal energy, the Bose-Einstein distribution for the phonon occupation can be approximated by linear temperature dependence: 
$N_{\omega_q}=1/(e^{\hbar\omega_q/k_BT}-1)\approx
k_BT/\hbar \omega_q, $
where $\omega_q$ is the frequency of a phonon with wavevector $q$.

\begin{figure}[t]
\centering
\includegraphics[width=\linewidth]{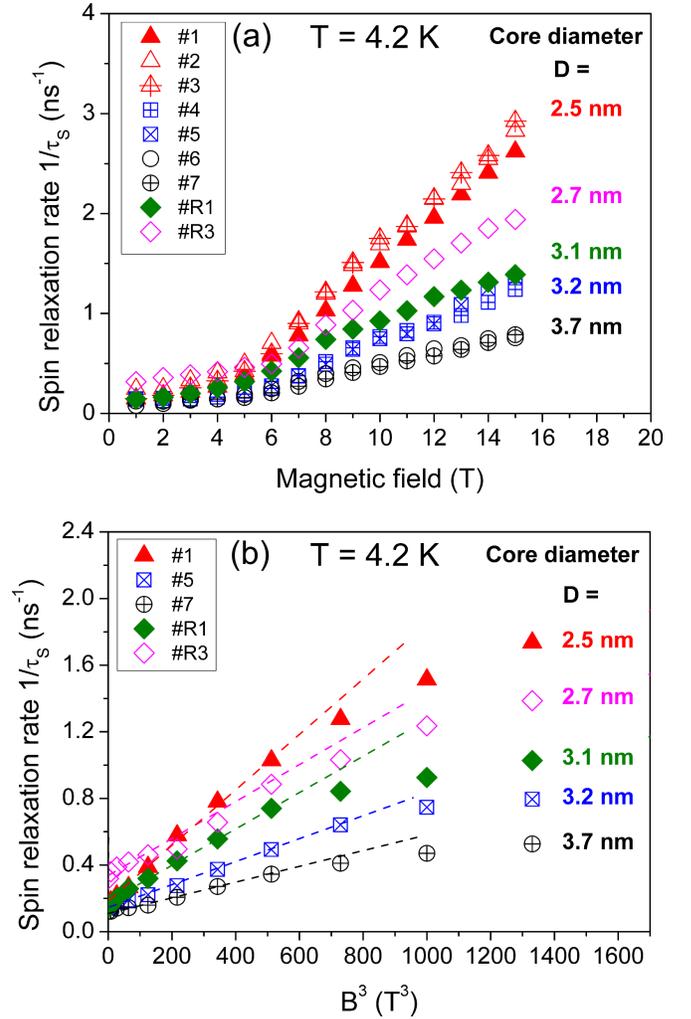}
\caption{(Color online) (a) Magnetic field dependence of the spin relaxation rate $1/\tau_s$ for the DiR samples \#1-\#7 and spherical core/shell samples \#R1 and \#R3 at $T=4.2$~K. (b) Magnetic field dependence of the spin relaxation rate $1/\tau_s$ for selected samples at low magnetic fields. Dashed lines are guides to the eye.}
\label{figure6}
\end{figure}

In order to get deeper insight into the origin of the spin relaxation we investigate its dependence on the NC shape. In Fig.~\ref{figure6}(a) the magnetic field dependences of the spin relaxation rate are shown for all studied DiR samples as well as the spherical samples \#R1 and \#R3. All of them show qualitatively the same trend, i.e. at low fields the spin relaxation rate follows a cubic dependence, as shown for selected samples in Fig.~\ref{figure6}(b), while at higher fields a linear magnetic field dependence is observed. However, the spin relaxation rate seems to be strongly dependent on the core size and independent of the shell shape. This is most obvious from the green and blue data points in Fig.~\ref{figure6}(a), which belong to the spherical NCs \#R1 and DiRs with a core size of $3.2$~nm. Although their shell shapes are totally different, the measured spin relaxation rates are quite similar.
This absence of shell shape dependence for the spin relaxation rate allows us to draw interesting conclusions to be discussed in the following.

The sensitivity of the spin relaxation rate on temperature suggests an acoustic phonon mediated process. Due to the increasing number of available phonons with increasing temperature the spin relaxation becomes more efficient. However, in contrast to bulk material, the phonon density of states in NCs is modified by confinement, as it has been confirmed in many experiments.\cite{Woggon1996,Saviot1996,Oron2009} This leads to the
''acoustic phonon bottleneck'',\cite{Fernee2012} meaning that low-energy, long-wavelength acoustic phonons are suppressed in nanometer-sized structures. Due to the similar elastic properties of CdSe and CdS it is expected that the phonon confinement is different in DiRs compared to spherical NCs. Namely the lower confinement along the rod axis presumably gives rise to longitudinal modes in addition to the torsional and spheroidal eigenmodes observed in spherical structures.\cite{Oron2009} This raises the question whether the phonon modes, which couple to the
excitons, are considerably different in DiR structures compared to the spherical case. In this respect we infer from the data presented in
Fig.~\ref{figure6}(a) that if acoustic phonons play a crucial role regarding the efficiency of the spin relaxation mechanism, the shell shape does not lead to a significantly changed number of phonon modes, which interact with the exciton. This conclusion is in agreement with the results of a recent experimental study employing fluorescent line narrowing (FLN) on DiRs in magnetic field. \cite{Aguila2014} Therein the authors found that the acoustic phonon modes of DiR nanostructures involved in exciton recombination in presence of a magnetic field are similar to those of confined modes in an isolated spherical NC.
Following the preceding line of thought, the magnetic field dependence of the spin relaxation rate could partly be explained in terms of the phonon bottleneck. For all studied samples the magnetic field dependence can be divided into two regimes. In the low field regime the relaxation rate follows a cubic dependence, whereas at higher fields the dependence becomes linear. On the other hand one can claim that at higher fields the slope of the spin relaxation rate dependence is increased compared to lower fields. The onset of this increase occurs at $B=8-10$~T. Assuming an average angle between the magnetic field and the crystal
c-axis of $45^{\circ}$ and taking $g_{ex}=2.7$ for the dark exciton $g$ factor,\cite{Biadala2010a} the corresponding Zeeman splitting amounts to $\Delta E \approx 1$~meV. This energy matches roughly the vibrational energy of the $l=2$ spheroidal mode of a CdSe sphere with a diameter of $5$~nm as estimated according to Lamb's theory.\cite{Oron2009} We propose that the observed change in the magnetic field dependence results from the onset of these spheroidal vibrations at magnetic fields around $8-10$~T. Although the core diameter of the samples studied here is considerably smaller than $5$~nm, we think that due to the similar elastic properties of CdSe and CdS, the overall size of the shell might be relevant in this regard.\cite{Liu2013} In the Lamb theory the energy of the spheroidal modes scales as $E\propto 1/D$, where $D$ is the sphere diameter. We therefore expect, that for larger NCs the onset of the linear magnetic field dependence should occur at lower fields. However, we are not able to verify such a trend from our data. Nevertheless we would like to stress, that our interpretation is in accordance
with the findings of a recent magneto-optical experiment on trions in CdSe/CdS core/shell NCs, where the spin relaxation was found to be completely inhibited for small Zeeman splitting, as a result of the acoustic phonon confinement.\cite{Fernee2012}

\begin{figure}[t!]
\centering
\includegraphics[width=\linewidth]{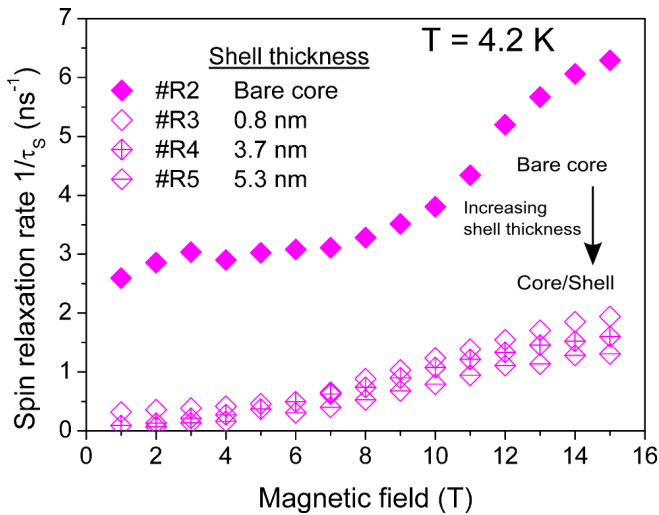}
\caption{(Color online) Magnetic field dependence of the spin relaxation rate  $1/\tau_s$ for spherical NCs \#R2-\#R5 with different shell thicknesses and similar core size of 2.7~nm.}
\label{figure7}
\end{figure}

The red data points in Fig.~\ref{figure6}(a), which belong to DiR samples with core diameter $2.5$~nm, indicate that the shell thickness might have a small effect on the spin relaxation rate. With increasing shell thickness the spin relaxation rate decreases slightly at a given magnetic field. In order to investigate this effect more systematically, we studied the set of spherical NCs \#R2-\#R5, which have similar core sizes but different shell thicknesses. Sample \#R2 consists of bare CdSe cores. The corresponding magnetic field dependences of the spin relaxation rates are shown in Fig.~\ref{figure7}. The most striking observation is that the rate of exciton spin relaxation in the bare core sample \#R2 is approximately one order of magnitude largeer (the spin relaxation time is as short as $\tau_s\approx 300$~ps already at low magnetic fields) compared to core/shell samples ($\tau_s \approx 3$~ns at low fields). The addition of a CdS shell leads to a strong drop of the spin relaxation rate, but further increase of the shell thickness from $0.8$~nm up to $5.3$~nm reduces the rate only slightly. This weak dependence on shell thickness was also observed in a previous magnetic field experiment on spherical core/shell ensembles. \cite{Liu2013} Here, only slight changes in the spin relaxation rate for shell thicknesses below $5$~nm were observed, whereas for even thicker shells the decrease got more pronounced. This was attributed to the onset of NC charging at shell thicknesses around $5$~nm, leading to very different spin dynamics of negatively charged excitons (trions) compared to the neutral excitons.

\begin{figure}[t!]
\centering
\includegraphics[width=\linewidth]{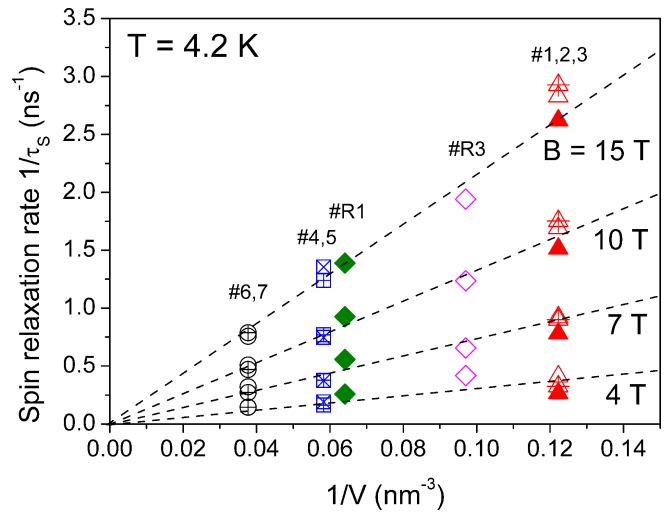}
\caption{(Color online) Dependence of the spin relaxation rate $1/\tau_s$ on core volume $V$ at different magnetic fields for the samples shown in Fig.~\ref{figure6}(a). Dashed lines are guides to the eye.}
\label{figure8}
\end{figure}

In Fig.~\ref{figure8} we show the spin relaxation rate as a function of core volume for the same samples as in Fig.~\ref{figure6}(a). Interestingly, the relaxation rate seems to be inversely proportional to the CdSe core volume. This observation, along with the surprisingly fast spin relaxation for bare cores and the weak dependence of $\tau_s$ on the shell thickness, substantiates the importance of the NC surface and the
CdSe/CdS interface in core/shell structures with respect to the spin relaxation mechanisms. We propose, that the spin relaxation rate depends on the strength of the interaction of the exciton with the surface or the interface. As the volume of the core increases, the surface to volume ratio of the core decreases and the interaction of excitons with the bare core surface or core/shell interface is continuously reduced.

It is well established that, due to the high valence band offset between CdS and CdSe the hole is confined to the core, whereas the electron wavefunction, due to the smaller conduction band offset, can leak into the shell to a certain degree.\cite{Brovelli2011} On the one hand this could explain the weak dependence of the spin relaxation rate on the shell thickness, since the overall surface in core/shell NCs is experienced only by the electron. On the other hand, this aspect and the strong core size dependence of $\tau_s$ indicate that the hole might be the important carrier regarding the exciton spin relaxation process.

It is worth noting, that changing the core volume or the shell thickness of the NC always results in a modification of the electron-hole exchange interaction due to the modified carrier wavefunction overlap. Thereby the exciton fine structure, and in particular the bright-dark energy splitting, are modified. This influences the mixing of bright and dark exciton states due to the perpendicular component of the external magnetic field. Nevertheless, considering the wide range of dark-bright energy splittings, which for the DiR samples investigated here vary approximately between 2 and 5~meV,\cite{Biadala2014} and considering the small change of the spin relaxation rate with shell thickness, we are convinced that the impact of changes of the exciton fine structure due to a varying electron-hole exchange interaction on the spin dynamics is small.

The apparent relevance of the core/shell interface for the spin relaxation is interesting, because its influence on the optical properties of NCs is yet poorly understood but undoubtedly important. For example, it was shown in Ref.~\onlinecite{Wang2009}, that a ''smooth transition'' from the core to the shell material favors non-blinking  emission of the NCs.

In this context, we want to highlight the theoretical studies of
Knipp and Reinecke as well as related work dealing with the interaction between charge carriers and acoustic phonons in semiconductor nanostructures,\cite{Reinecke1995,Knipp1996,Alcalde2000,Alcalde2002} which to the best of our knowledge have not been considered explicitly in relation with colloidal NCs. In these studies the authors investigate a new type of electron-phonon coupling, which they name the ''ripple mechanism''. This coupling is related to the spatial dependence of the electron effective mass and its potential energy. The coupling occurs due to the motion of the interface between the nanostructure and the surrounding medium. Notably, the authors predict this effect to be dominant over the common deformation potential coupling for small NCs, as studied here. Moreover they expect the phonon scattering rate to be strongly dependent on the nanostructure size with higher scattering rates for smaller NCs. As it was discussed for epitaxially grown III-V quantum dots, the ripple mechanism also strongly affects the electron \cite{Woods2002} and hole \cite{Woods2004} spin relaxation in magnetic field.

The coupling between excitons and phonons in colloidal NCs is a topic, which is still poorly understood, although it obviously plays an important role for energy and spin relaxation as well as the recombination of excitons. The aforementioned studies stress the relevance of the interface between the nanostructure and its surrounding medium regarding exciton-phonon coupling as well as spin relaxation and our experimental results point in the same direction. However, we think, that exciton spin relaxation in NCs might be result of a complicated interplay of many effects, since for example the presence of magnetic moments of defects or dangling bonds, which are often considered in the context of surface or interface related phenomena, have not been explicitly considered here.

\subsection{Equilibrium circular polarization degree}

\begin{figure}[h]
\centering
\includegraphics[width=\linewidth]{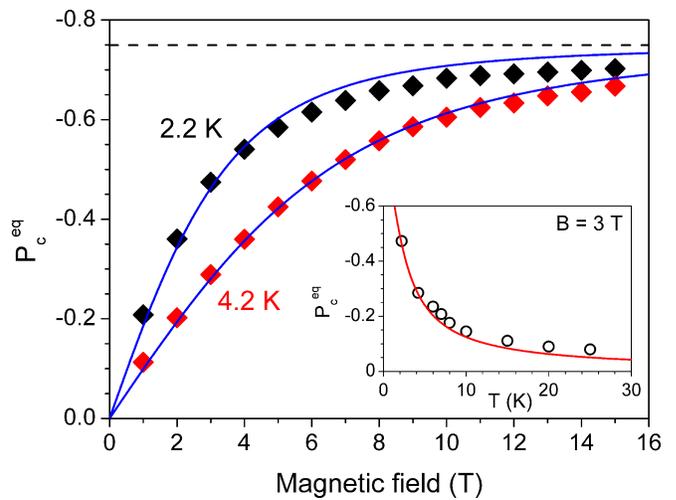}
\caption{{(Color online) Magnetic field dependence of $P_c^{\text{eq}}$ for sample \#R1 at $T=2.2$~K and $4.2$~K. The black dashed line indicates the theoretical limit of $P_c^{\text{eq}}=-0.75$ for an ensemble of randomly oriented NCs. The blue solid lines are fits of Eq.~(\ref{bdependence}) to the experimental values, assuming $f_{\text{or}}(x)=1$. The inset shows the temperature dependence of $P_c^{\text{eq}}$ at $B=3$~T. The red line is calculated taking an exciton $g$ factor $g_{ex}=2.5$ as determined from the fit of the magnetic field dependence.}}
\label{figure9}
\end{figure}

In this section we discuss the dependence of $P_c^{\text{eq}}$ derived from the fit of the time-resolved DCP curves with
Eq.~(\ref{fit}) on magnetic field, temperature and NC shape. In
Fig.~\ref{figure9} the magnetic field dependence of
$P_c^{\text{eq}}$ is shown for the spherical core/shell sample \#R1 at $T=2.2$~K and 4.2~K. At low magnetic fields $P_c^{\text{eq}}$ shows a steep increase, whereas at high magnetic fields it saturates. The blue solid lines are fits of the experimental data with Eq.~(\ref{bdependence}), assuming a randomly orientated NC ensemble, i.e. $f_{\text{or}}(x)=1$. In this case, the theoretical limit for $P_c^{\text{eq}}$ is expected to be $-0.75$, as indicated by the black dashed line. The fit yields an exciton $g$ factor of $g_{ex}=2.5$, which is close to the recently reported values for the dark exciton $g$ factor of $g_{ex}=2.8$ in a FLN experiment\cite{Aguila2014} and $g_{ex}=2.7$ in a single-NC PL experiment\cite{Biadala2010a}, although similar values of
$g_{ex}=2.4-2.6$ were also reported for the bright exciton.\cite{Htoon2009} At a given magnetic field $P_c^{\text{eq}}$ decreases with increasing temperature, see the inset of Fig.~\ref{figure9}. An increase of the temperature from $2.2$~K up to $25$~K leads to a drop of $|P_c^{\text{eq}}|$ from approximately $0.5$ down to $0.1$. This is result of thermal population of the upper Zeeman-sublevel, corresponding to a reduced exciton spin polarization. The experimental data can be fitted well with Eq.~\eqref{bdependence} using $g_{ex}=2.5$, see the red line.

\begin{figure}[t]
\centering
\includegraphics[width=\linewidth]{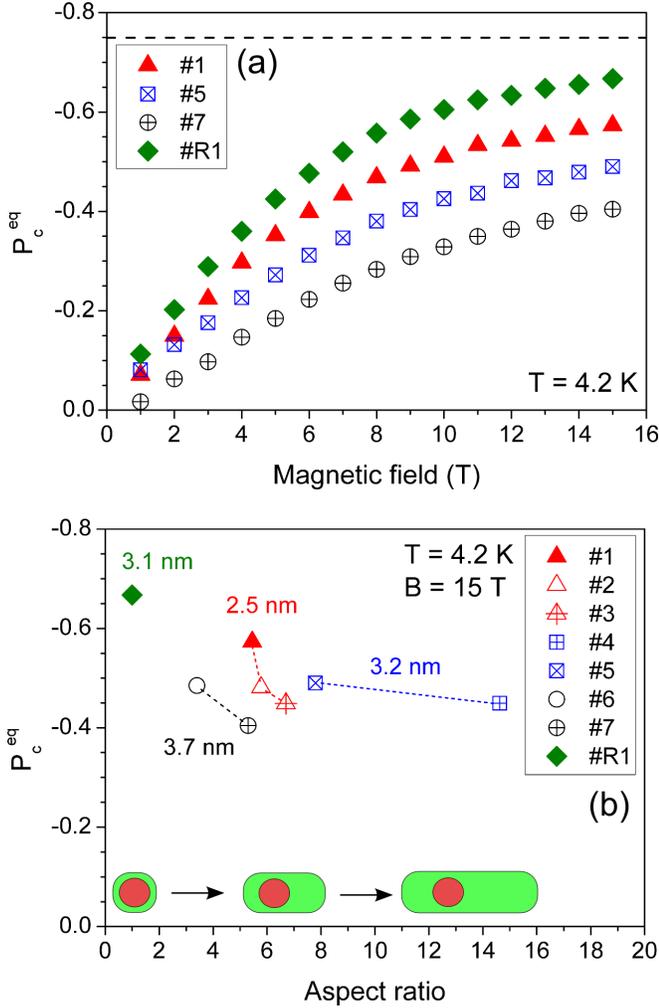}
\caption{{(Color online) (a) Magnetic field dependences of $P_c^{\text{eq}}$ for samples of different core size at $T=4.2$~K. The black dashed line indicates the theoretical limit of $P_c^{\text{eq}}=-0.75$ for an ensemble of randomly oriented NCs. (b) $P_c^{\text{eq}}$ for the DiR samples \#1-\#7 and the spherical core/shell sample \#R1 at $T=4.2$~K  and $B=15$~T as function of the aspect ratio. The inset schematically illustrates the corresponding change of NC shape. The dashed lines link data points for NCs with the same core sizes and are guides to the eye.}}
\label{figure10}
\end{figure}

In Fig.~\ref{figure10}(a) we show $P_c^{\text{eq}}$ as a function of
$B$ for several DiR samples together with the spherical core/shell sample \#R1 at $T=4.2$~K. Apparently, the evolution of $P_c^{\text{eq}}$ with $B$ is quite different for samples of different shape and different core size and we always find a lower polarization degree for DiR NCs. The dependences differ in their slope at low magnetic fields and also appear to have very different saturation levels, although saturation is not fully reached within the magnetic field range of our experiment. A varying slope at low magnetic fields could result from differences of the
exciton $g$ factor, which is mainly determined by the variation of the electron $g$ factor. In the studied structures the holes are strongly localized in the core and the hole $g$ factor can be estimated using the expression developed for a potential of spherical symmetry as $g_h=-1.09$, see Refs.\onlinecite{Gelmont1973,Efros1996}. It is independent of the core size and is therefore expected to be the same for all samples. However, the increase of the CdS shell leads to a decrease of the electron quantum confinement energy and thus to a decrease of the electron $g$ factor because of its energy dependence.\cite{Rodina2003} Simultaneously, the leakage of the electron wave function into the  CdS shell leads to a contribution from the CdS $g$ factor, which is close to 2, and thus to an increase of the electron $g$ factor. These two effects may partly compensate each other. However, the latter can even result in an axial anisotropy of the electron $g$ factor.

Nevertheless, differences in the $g$ factor do not affect the saturation level of $P_c^{\text{eq}}$ at high magnetic fields. Although saturation is not reached in the magnetic field range of our experiment, the data in Fig.~\ref{figure10}(a) indicate saturation of $P_c^{\text{eq}}$ at different levels. In Fig.~\ref{figure10}(b) $P_c^{\text{eq}}(B=15$~T) is shown for all DiR samples as well as sample \#R1 as function of the aspect ratio (the ratio of length to width of the NC shell given in Table~\ref{table1}). The values of $P_c^{\text{eq}}(B=15$~T) vary between $-0.66$ for the spherical sample \#R1 and $-0.4$ for DiR sample \#7. For a given core size, we find, that $P_c^{\text{eq}}$ is always lower for the NCs with higher aspect ratio. This observation indicates, that the differences in $P_c^{\text{eq}}$ could be related to the geometry of the NC shell. As discussed in Ref.~\onlinecite{Yakovlev2002}, the circular polarization of the PL in magnetic field can be reduced by an optical anisotropy, resulting from linear polarization of the individual emitters. Typical values for the PL linear polarization along the anisotropic c-axis of individual DiRs were reported to be close to 75\% at room temperature.\cite{Talapin2003,Pisanello2010,Hadar2013} To the best of our knowledge, there are no reports about the linear polarization for single DiR NCs at low temperatures. The high linear polarization degree of the PL observed for single DiR NCs at room temperature may result partly from the exciton fine structure and partly from a dielectric enhancement effect,\cite{Shabaev2004,Diroll2014} both of which are dependent on the NC geometry. In the following section we provide a detailed discussion of these effects and analyze how they affect the circular and linear polarization of the PL in magnetic field for NCs with various shapes at low temperatures.

\section{Theoretical treatment of linear and circular polarization of
low temperature PL in external magnetic field} \label{theorypart}
\subsection{Model}

\begin{figure}[t!]
\centering
\includegraphics[width=\linewidth]{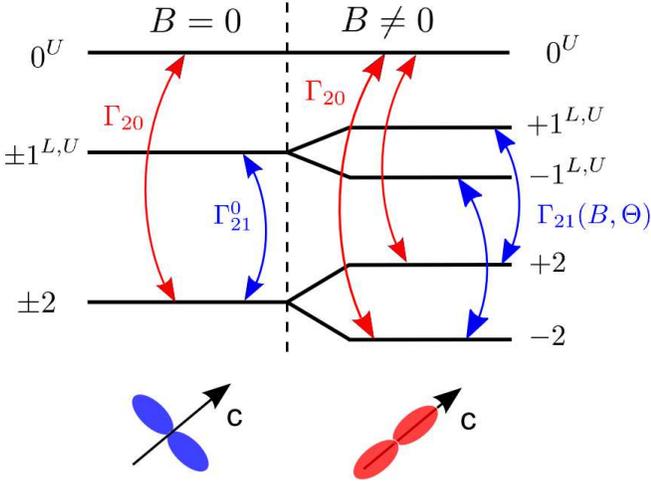}
\caption{{(Color online) Schematic of the band edge exciton energy levels in zero and nonzero magnetic fields. Only the dark exciton ground state with $F=\pm 2$ and the bright exciton states $F=\pm 1^{L,U}$ and $F=0^U$ are shown (without splitting between the $1^L$ and $1^U$ states). The red and blue arrows show the coupling pathways, which may enable dark exciton radiative recombination. The sketches at the bottom illustrate the respective orientation of the transition dipole for the different coupling pathways with respect to the crystal c-axis. Here colors correspond to those of the arrows indicating the coupling pathways.}}
\label{figure11}
\end{figure}

\begin{figure}[t!]
\centering
\includegraphics[width=\linewidth]{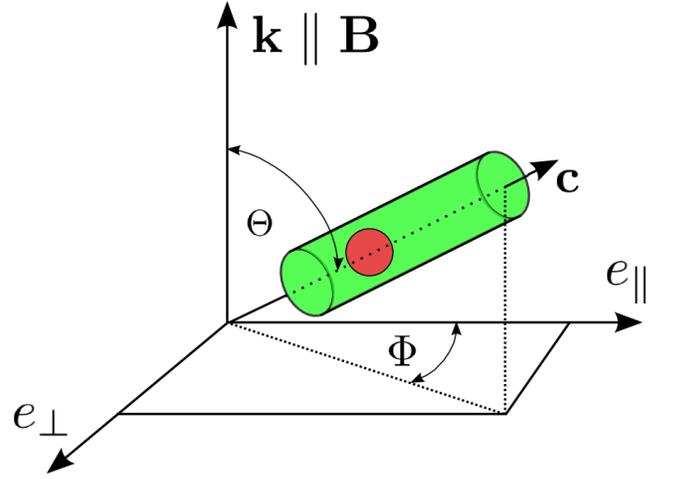}
\caption{{(Color online) Schematic of the experimental geometry for measuring circular and linear polarization of the PL from DiRs or spherical NCs with anisotropic c-axis. Note that in the ensemble the angles $\Theta$ and $\Phi$, which determine the direction of $\mathbf{c}$ with respect to the light vector $\mathbf{k}$ and the linear polarizer axis $\mathbf{e}_\parallel$, are arbitrary.}}
\label{figure12}
\end{figure}

We start with the assumption, that the lowest exciton state is the dark exciton with momentum projection $|F|=2$ on the hexagonal c-axis (which is also the rod axis in DiR NCs). Figure~\ref{figure11} shows a sketch of the exciton energy levels in zero and nonzero magnetic field. Only the ground state $F=\pm 2$ and the bright exciton states $F=\pm 1^{L,U}$ and $F=0^U$ are shown (without the energy splitting between the $1^L$ and $1^U$ states). Furthermore, we assume that the dark exciton $|F|=2$ can recombine radiatively only via coupling with the bright exciton states. Coupling with the $|F|=1^{L,U}$ states leads to  recombination which polarization is oriented isotropically in a plane perpendicular to $c$-axis.\cite{Empedocles1999} The respective recombination rate is denoted by $\Gamma_{21}^0$ at zero and $\Gamma_{21}$ at nonzero magnetic field so that the probability to emit light in the direction ${\bm k}$ (see the schematic of the experimental geometry Fig. 12) is given by
$\Gamma_{21}^0(1+\cos^2 \Theta)$ and   $\Gamma_{21}(1+\cos^2 \Theta)$, respectively.
Moreover the coupling with the $|F|=1$ states can be enhanced by a magnetic field oriented perpendicular to the c-axis.\cite{Efros1996,Johnston-Halperin2001} For this reason, $\Gamma_{21}$ depends on the angle $\Theta$ between the magnetic field direction and the c-axis (see
Fig.~\ref{figure12}).
In contrast, the coupling of the dark exciton $|F| = 2$ with the bright $F=0^U$ state leads to recombination polarized parallel to the c-axis.  Its recombination rate, which is
denoted by $\Gamma_{20}$, is independent of the magnetic field and the probability to emit light in the direction ${\bm k}$  is given by $\Gamma_{20}\sin^2 \Theta$. The total recombination rate of the dark exciton averaged over all light directions is proportional to $\Gamma_{20}+2\Gamma_{21}$.

We consider hereafter only the range of magnetic field satisfying the condition $g_{ex}\mu_B B^{\text{sat}} \ll \Delta E_{bd}$. In this range the recombination rates of $F=-2$ and $F=+2$ excitons calculated in the framework of second order perturbation theory  are equal.

It should be noted that the usual assumption leading to Eq.~(\ref{bdependence}) for describing the magnetic field dependence of $P_c^{eq}$ is that the dark exciton recombines due to coupling with the $|F|=1^{L,U}$ bright states\cite{Efros1996,Johnston-Halperin2001}, characterized by $\Gamma_{21}$ and thus has the same polarization properties as the $|F|=1$ exciton. In this case, the dipole moment acquired by the dark exciton is oriented transverse to the c-axis, see the scheme left bottom in Fig.~\ref{figure11}. The new aspect of our consideration is the assumption, that the dark exciton can recombine with the properties of the $F=0$ bright exciton as well. This coupling between the $|F|=2$ dark exciton and the $F=0$ state described by $\Gamma_{20}$ could be mediated by $l=2, m=\pm 2$ acoustic phonons
\cite{Chamarro1996,RodinaUnpublished}. It could be also caused by the shape anisotropy in the plane perpendicular to the c-axis, which mixes the valence band states \cite{Goupalov2006} but more importantly leads to an anisotropy of the electron wavefunction leaking into the rod and thus to an anisotropy of the Coulomb interaction in the exciton.

Importantly, in DiR NCs the recombination polarized along the c-axis can be additionally enhanced by the dielectric enhancement effect. The enhancement factor $R_e$ in Ref.~\onlinecite{Shabaev2004} is given by
\begin{equation}
R_e=\left(\frac{\kappa_m+(\kappa_s-\kappa_m)n^{(x,y)}}{\kappa_m+(\kappa_s-\kappa_m)n^{(z)}}\right)^2,
\label{revalues}
\end{equation}
where $\kappa_m$ and $\kappa_s$ are the dielectric constants of the surrounding medium and the semiconductor, respectively.
$n^{(x,y)}=(1-n^{(z)})/2$ and $n^{(z)}=a^2[\ln{(2b/a)}-1]/b^2$ are the depolarization coefficients along the directions perpendicular (x,y) and along (z) the hexagonal axis. The NCs are approximated by ellipsoids with major semi-axis $b$ and minor semi-axis $a$. Essentially, the rod shaped shell of DiRs acts as a linear polarizer, which affects the absorption as well as the emission of light. Note that here we neglect the frequency dependence of $R_e$ and assume that it is the same for excitation and emission.

The dielectric enhancement effect can lead to selective excitation of DiR
NCs oriented perpendicular to the magnetic field (the light propagation direction). According to Ref.~\onlinecite{Kovalev1996}, the probability for a NC to be excited by linearly polarized light is
$1+(R_e-1)(\mathbf{e}\cdot \mathbf{c})^2$, where $\mathbf{e}$ is the light polarization vector. For the geometry shown in Fig.~\ref{figure12} and light polarized along the $e_{\parallel}$ direction one can write $(\mathbf{e}\cdot \mathbf{c})^2=\sin^2\Theta\cos^2\Phi$. By averaging over all polarizations in the plane $(e_{\parallel},e_{\perp})$ we obtain the probability for excitation of a NC, oriented at an angle $\Theta$ with respect to the light propagation direction, by unpolarized light as
\begin{equation}
f_\text{exc}(x)=1+0.5(R_e-1)(1-x^2) \, ,
\end{equation}
where $x=\cos\Theta$. Taking into account the selective excitation effect, the general expression Eq.~(\ref{bdependence_gen}), which describes the DCP for an ensemble of DiR NCs, transforms into
\begin{equation}
P_c(B)=\frac{\int_0^1
[I_{\sigma+}(x)-I_{\sigma-}(x)]f_\text{or}(x)f_\text{exc}(x)dx}{\int_0^1
[I_{\sigma+}(x)-I_{\sigma-}(x)]f_\text{or}(x)f_\text{exc}(x)dx}.
\end{equation}
In the following we consider only the case of randomly oriented NCs,
i.e. $f_\text{or}(x) \equiv 1$. When the selective excitation is taken into account and the $\Gamma_{20}$ rate is neglected, expression Eq.~(\ref{bdependence}) for the magnetic field dependence of the DCP remains unchanged with the function $f_\text{or}(x)$ replaced by $f_\text{exc}(x)$.

When we additionally consider the activation of the dark exciton by a magnetic field applied perpendicular to the c-axis, which leads to a  field and angle dependence of the rate $\Gamma_{21}(B,x)$, Eq.~\eqref{bdependence} transforms into
\begin{equation}
P_c^{\text{eq}}(B) =\frac{\int_0^1 \gamma(B,x)2x\tanh(\Delta E
/2k_BT)f_\text{exc}(x) dx}{\int_0^1
\gamma(B,x)(1+x^2)f_\text{exc}(x) dx} \, , \label{bdependence_eq1}
\end{equation}
where the factor $\gamma(B,x)=\Gamma_{21}(B,x)/\Gamma_{21}^0$ is a measure of the strength of the magnetic-field-induced mixing.

In general, the angular and magnetic field dependence of $\Gamma_{21}(B,x)$ can vary, depending on the ratio between
$g_{ex}\mu_B B^{\text{sat}}$ and the bright-dark exciton splitting
$\Delta E_{bd}$. $B^{\text{sat}}$ is the magnetic field at which the DCP is saturated.
For the range of magnetic fields we consider here the condition $g_{ex}\mu_B B^{\text{sat}} \ll \Delta E_{bd}$ is satisfied and
one can write $\Gamma_{21}(x)=\Gamma_{21}^0[1+p(1-x^2)]$,
where the parameter $p$ describes the dependence on $B$ with $p \propto B^2$.\cite{Efros1996,Efros2003}

Now we also include the coupling of the $|F|=2$ states to the state
$F=0^U$. The transitions polarized along the c-axis contribute equally to $I_{\sigma+}(x)$ and $I_{\sigma-}(x)$ with the probability $\Gamma_{20} R_e\sin^2 \Theta $. Accounting for these transitions we obtain the following expression for an ensemble of randomly oriented DiR NCs:
\begin{equation}
P_c^{\text{eq}}(B) =\frac{\int_0^1 \gamma(B,x)2x\tanh(\Delta E
/2k_BT)f_\text{exc}(x) dx}{\int_0^1
\left[\gamma(B,x)(1+x^2)+r(1-x^2)\right] f_\text{exc}(x) dx} \, ,
\label{bdependence_eq}
\end{equation}
where $r=R_e\Gamma_{20}/\Gamma_{21}^0$. Obviously, on the one hand the coupling to the $F=0^U$ state lowers $P_c^{\text{eq}}$ at a given magnetic field. On the other hand it leads to a higher degree of linear polarization for the individual NC, as will be discussed below.

The degree of linear polarization for an individual NC, is $\rho_{l}=(I_\parallel -I_\perp)/(I_\parallel
+I_\perp)$. Here $I_\parallel(I_\perp)$ is the intensity of light
polarized along the $e_\parallel(e_\perp)$ direction.  It can be shown that
\begin{equation}
\rho_l(x,\Phi)=\frac{(1-x^2)(R_e\Gamma_{20}-\Gamma_{21})\cos(2\Phi)}
{(1+x^2)\Gamma_{21}+(1-x^2)R_e\Gamma_{20}} \, .
\end{equation}
One sees that after averaging over the randomly oriented ensemble of DiR NCs, the overall linear polarization $P_l=\int_0^1dx\int_0^{2\pi}\rho_l(x,\Phi)d\Phi$ is zero. However, it still can be nonzero for the individual DiR NC reaching a maximum for rods oriented at $\Phi=0^{\circ}$ and $\Theta=90^{\circ}$ (i.e. $x=0$). In the following we consider only the maximum linear polarization for this particular geometry which for zero magnetic field gives
\begin{equation}
\rho_{l,0}^{\text{max}}=\frac{r-1}{r+1} \, .
\end{equation}
It is worth noting, that for the case $r<1$ (i.e. a higher coupling rate to the $\pm 1$ states $\Gamma_{21}^0$ in comparison to the dielectrically enhanced rate $R_e\Gamma_{20}$), $\rho_{l,0}^{\text{max}}$ has negative values. This corresponds to the case, where the light is preferentially polarized perpendicular to the c-axis. In magnetic fields at which the circular polarization degree is saturated ($B=B^{\text{sat}}$):
\begin{equation}
\rho_l^{\text{max}}(B^{\text{sat}})=\frac{r-(1+p^{\text{sat}})}{r+(1+p^{\text{sat}})}
\, .
\end{equation}
Here we have defined
$(1+p^{\text{sat}})=\gamma(B^{\text{sat}},x=0)=\Gamma_{21}(B^{\text{sat}},x=0)/\Gamma_{21}^0$. It is a measure for the increase of $\Gamma_{21}$ at $B=B^{\text{sat}}$ for NCs oriented perpendicular to the magnetic
field.

So far, the equations for the circular polarization of the DiR ensemble and the linear polarization of individual DiRs in magnetic field have been derived.  We turn now to the analysis of the correlation between the linear polarization of individual DiRs and the ensemble DCP in more detail and compare our theory to the experimental data on DiRs.

\subsection{Discussion}

In the following, we analyze how the saturation level of
$P_c^{\text{eq}}$ (i.e. $P_c^{\text{sat}}$) is affected by the different mechanisms we introduced in the previous section. To that end, we simplify the discussion by analyzing first the effect of selective excitation and the newly introduced coupling to the $F=0^U$ state, while we set $p=p^{\text{sat}}=0$, which means that magnetic field activation of dark exciton radiative recombination is neglected as first approximation. Later on the additional effect of $p\neq 0$ is discussed.

\begin{figure*}[t!]
\centering
\includegraphics[width=\linewidth]{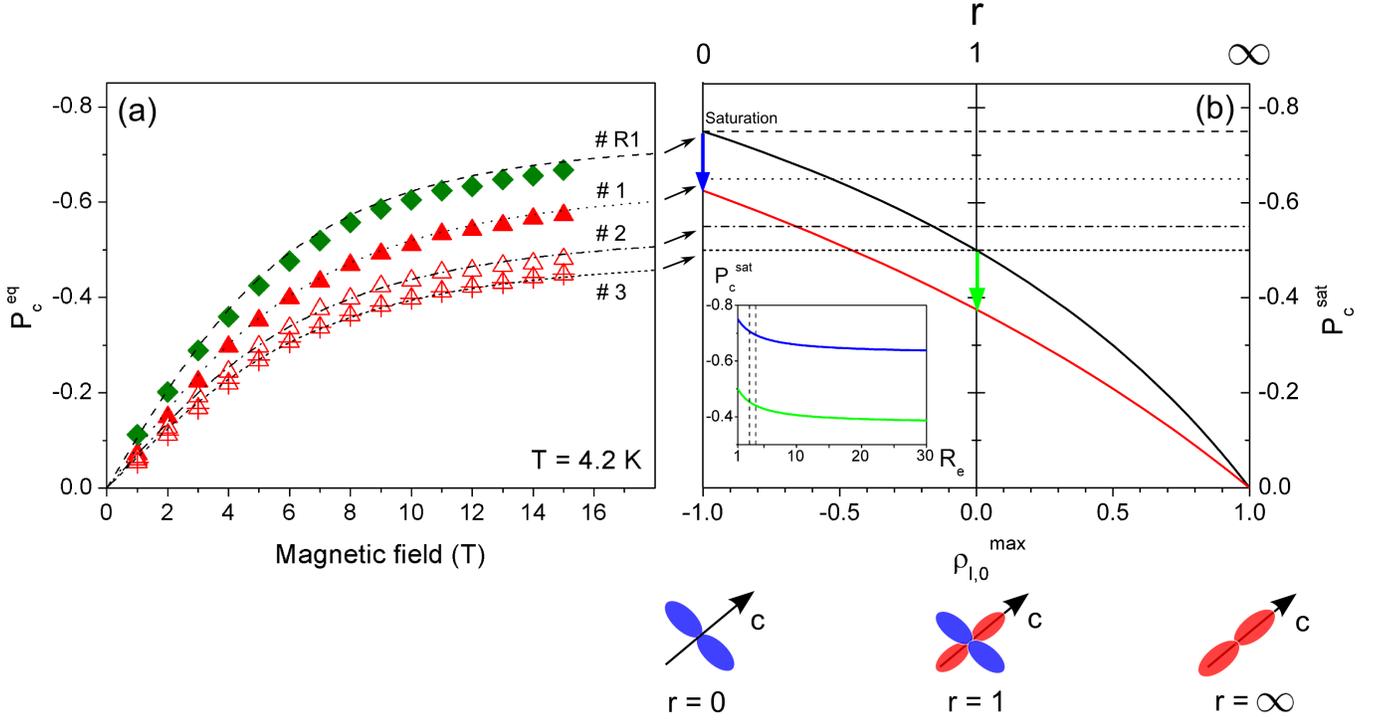}
\caption{{(Color online) (a) Magnetic field dependences of $P_c^{\text{eq}}$ for the DiR samples \#1, \#2, \#3 and the spherical core/shell sample \#R1 at $T=4.2$~K. The dashed and dotted lines are calculated according to Eq.\eqref{bdependence_eq}. (b) Saturation values of the circular polarization in high magnetic fields $P_c^{\text{sat}}$ for an ensemble of arbitrarily oriented NCs vs. maximal linear polarization of individual NCs in zero magnetic field $\rho_{l,0}^{\text{max}}$. The orientation of the transition dipoles with respect to the NC c-axis is illustrated in the sketches below. The black curve is calculated without considering the effect of selective excitation (i.e. $R_e=1$) and the red curve is calculated taking $R_e=1000$, which corresponds to an extremely strong effect of selective excitation. The inset shows $P_c^{\text{sat}}$ as a function of $R_e$ for $\rho_{l,0}^{\text{max}}=-1$ (blue curve) and $\rho_{l,0}^{\text{max}}= 0$ (green curve). The dashed lines in the inset mark the range of calculated enhancement factors for the investigated samples according to Eq.~\eqref{revalues}. The dashed and dotted lines in the main plot of panel (b) are the saturation values derived from the corresponding calculated curves in panel (a). All calculations were done neglecting the effect of magnetic field activation of the dark exciton recombination.}}
\label{figure13}
\end{figure*}

In Fig.~\ref{figure13}(a) we show the magnetic field dependences of $P_c^{\text{eq}}$ for the DiR samples \#1-\#3 and the spherical core/shell sample \#R1 at $T=4.2$~K. Obviously, the dependences are quite different and the $P_c^{\text{eq}}$ tend to saturate at different values. The lines are calculations according to Eq.~(\ref{bdependence_eq}). The parameters used are $g_{ex}=2.7$,   $R_e=3.2, 3.3, 3.4, 1$ and $\Gamma_{20}/\Gamma_{21}^0=0.03,0.13,0.19,0.02$ for samples \#1,\#2,\#3 and \#R1, respectively. This means that we assume the ratio between the coupling of the dark exciton to the $F=0$ state and the coupling to the $F=\pm 1$ states to be different for the different NCs.  The enhancement factors $R_e$ were calculated according to Eq.~(\ref{revalues}), where the dielectric constants where taken to be $\kappa_m=2$ and $\kappa_s=6$ for the surrounding medium and the NC, respectively.\cite{Shabaev2004} Apparently, the experimentally measured $P_c^{\text{eq}}$ can be well reproduced by our model.

Figure~\ref{figure13}(b) shows a plot of the saturated ensemble polarization $P_c^{\text{sat}}$ (for randomly oriented DiRs) as function of the maximal linear polarization of individual DiRs at zero magnetic field $\rho_{l,0}^{\text{max}}$. The magnetic field activation of the dark exciton is still neglected (i.e. $p^{\text{sat}}=0$). The horizontal lines correspond to the saturation values of $P_c^{\text{eq}}$ derived from the corresponding calculated curves in the left panel (the black arrows indicate saturation). The red and black curves are the theoretically predicted dependences of $P_c^{\text{sat}}$ on $\rho_{l,0}^{\text{max}}$ with and without considering the effect of selective excitation. Obviously, the selective excitation effect leads to an additional decrease of the saturated polarization value. The dependence of $P_c^{\text{sat}}$ on $R_e$ due to selective excitation is
shown in the inset for $\rho_{l,0}^{\text{max}}=-1$ and $\rho_{l,0}^{\text{max}}=0$ (see the corresponding blue and green arrows). The dashed lines indicate the range of enhancement factors calculated for the DiRs investigated here. Notably, even in the case of an extremely strong dielectric enhancement effect, the saturation value is not lowered by more than approximately $12$\%. In contrast, as can be seen from the red and black curves, the increase of the $r$ value (i.e. the ratio of the coupling rates, see the upper scale in Fig.~\ref{figure13}(b)) and the resulting maximal linear polarization of individual DiRs at zero magnetic field leads to a decrease of the saturation value of the circular polarization in the ensemble that can vary between $-0.75$ and $0$. From the intersection of the dashed and dotted lines (the theoretically derived saturation values) with the black curve, we can predict within this approximation the maximal linear polarization of the individual rod at zero field to be negative for all DiRs considered here ($\rho_{l,0}^{max}<0$ ). This means that we expect the light emitted from these DiRs to be linearly polarized predominantly transverse to the c-axis at low temperatures, see the sketches below Fig.~\ref{figure13}(b).

\begin{figure}[t!]
\centering
\includegraphics[width=\linewidth]{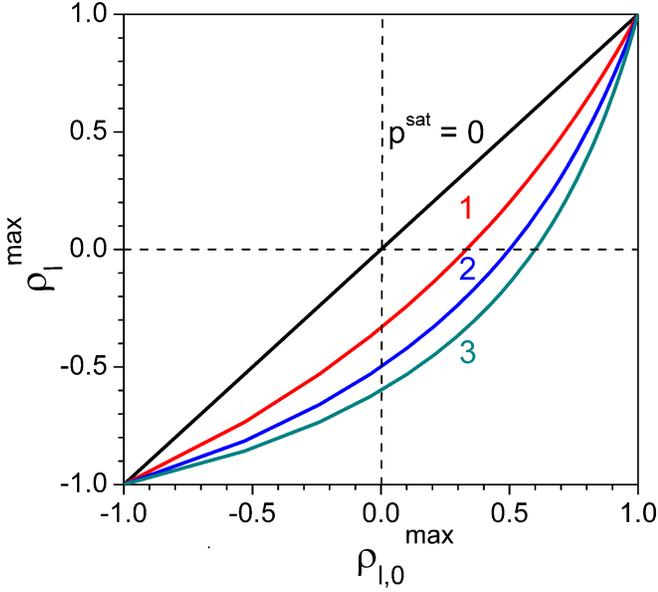}
\caption{{(Color online) Maximal values of individual NC linear polarization in high magnetic field $\rho_{l}^{\text{max}}$ vs. maximal values of the linear polarization in zero magnetic field $\rho_{l,0}^{\text{max}}$ for $p^{\text{sat}}=0,1,2,3$.}}
\label{figure14}
\end{figure}

Next, we discuss in more detail how the magnetic field activation of the dark exciton influences the polarization properties of the DiR emission. First, we analyze the effect on the linear polarization of individual rods. In Fig.~\ref{figure14} we plot the maximal linear polarization of individual rods at $B=B^{\text{sat}}$ versus the maximal linear polarization at zero magnetic field for different values of $p^{\text{sat}}$. We remind, that the parameter $p$ was assumed to be dependent on the magnetic field. Since we consider here the linear polarization either at $B=0$ or $B=B^{\text{sat}}$, we use
$p(B^{\text{sat}})\equiv p^{\text{sat}}=\gamma(B^{\text{sat}},0)-1$ as model parameter. It is defined by the coupling of the $F=\pm 2$ states to the $F=\pm 1$ states at zero and high magnetic fields and for a NC with the hexagonal axis oriented perpendicular to the magnetic field.
The magnetic field enhances the coupling to the $F =\pm 1$ states and therefore the amount of transitions polarized transverse to the c-axis increases with increasing magnetic field. This effect can be seen from the curves shown in Fig.~\ref{figure14}. If $p^{\text{sat}}=0$ the maximal linear polarization is independent of the magnetic field, i.e.
$\rho_{l}^{\text{max}}=\rho_{l,0}^{\text{max}}$. However, with increasing $p^{\text{sat}}$ the linear polarization in high magnetic fields shifts continuously to negative values, which means that the transitions polarized transverse to the c-axis are enhanced. This can lead to a situation in which at zero magnetic field the PL is polarized parallel to the c-axis, whereas at high fields it is polarized transverse to it.

The parameter $p^{\text{sat}}$ is unknown for the samples investigated here but could be determined by measuring the linear polarization of an individual DiR at zero and high magnetic fields. Such an experiment goes beyond the scope of this study. However, we can estimate the order of magnitude of $p^{\text{sat}}$ by considering the shortening of the PL decay typically observed in magnetic field experiments. In Ref.~\onlinecite{Biadala2010a}, which reports on magneto-optical PL measurements on single spherical CdSe/ZnS core/shell NCs, the authors observed a shortening of the PL decay from approximately $180$~ns to $50$~ns for a NC oriented at $75^\circ$ with respect to the magnetic field. This corresponds to a ratio of $3.6$ for the recombination rates and therefore $p$ is equal to 2.6. We expect a slightly larger shortening for larger orientation angles, thus the upper limit of
$p^{\text{sat}}$ should be around $3$. It is worth noting, that this approximation works only, if one assumes, that there is no contribution from the $F=0^U$ states to the PL, which can reasonably be assumed for NCs, where the DCP saturation value in a randomly oriented ensemble approaches $-0.75$ at high magnetic fields.

\begin{figure}[t!]
\centering
\includegraphics[width=\linewidth]{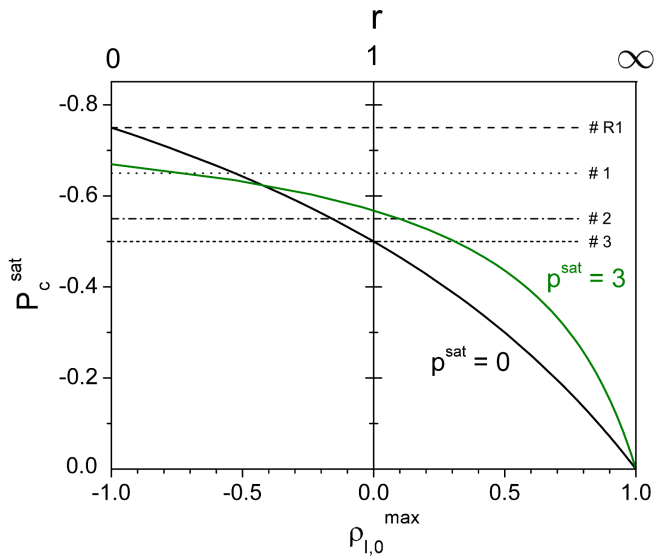}
\caption{{(Color online) Saturation values of the circular polarization degree in high magnetic fields $P_c^{\text{sat}}$ for an ensemble of arbitrarily oriented NCs vs maximal linear polarization values of individual NCs in high magnetic fields $\rho_{l,0}^{\text{max}}$, taking into account the effect of magnetic field activation of the dark exciton recombination ($p^{\text{sat}}=0$ for the black curve and $p^{\text{sat}}=3$ for the green curve). Calculations are done without considering the effect of selective excitation ($R_e=1$).}}
\label{figure15}
\end{figure}

Now, we analyze how the increased transverse linear polarization due to magnetic field activation of dark exciton radiative recombination affects the circular polarization of the ensemble PL. In
Fig.~\ref{figure15} we plot the saturation value of the ensemble circular polarization at $B=B^{\text{sat}}$ as function of the individual NC linear polarization at $B=0$ for $p^{\text{sat}}=0$ (black curve) and $p^{\text{sat}}=3$ (green curve). The selective excitation effect is not considered here (i.e. $R_e=1$). The horizontal lines are similar to those in Fig.~\ref{figure13}(b) and correspond to the predicted saturation levels for the samples shown in Fig.~\ref{figure13}(a). They are given here for better comparability. It is worth noting, that with accounting for the angular dependence of the dark exciton recombination rate, the predicted level of the saturateion DCP values from fitting the magnetic field dependences $P_c^{\text{eq}}(B)$ will be lower for each sample. One can see that in the region of small values of $r$ (see the upper scale of Fig.~\ref{figure15}), the effect of the dark exciton magnetic field activation leads to a decrease of $P_c^\text{sat}$ averaged over the ensemble. This is because the magnetic field activation depends on the angle and the resulting transitions become stronger for the NCs oriented perpendicular to the c-axis where the Zeeman splitting is smaller. With the increase of $r$, however, the increase of the $\Gamma_{21}$ transition becomes more important than the angular dependence and the resulting $P_c^\text{sat}$ is increased. As a result, from the intersection of the dashed and dotted lines with the green curve for $p^{\text{sat}}=3$, we can predict the maximal linear polarization of an individual rod at zero magnetic field. It is positive for the DiRs \#2 ($\rho_{l,0}^{\text{max}}=0.10$) and \#3 ($\rho_{l,0}^{\text{max}}=0.32$), so that we expect the light emitted from these DiRs to be linearly polarized predominantly along to the c-axis at low temperatures. However, as can be seen from
Fig.~\ref{figure14}, the maximal linear polarization of an individual rod at magnetic field $B=B^{\text{sat}}$ is negative: $\rho_{l}^{\text{max}}=-0.53$ and $-0.38$ for the samples \#2 and \#3, respectively.

As shown above, our model, which introduces the coupling of the dark exciton to the bright exciton state $F=0^U$ in addition to the well-known coupling to the $\pm 1$ states, allows one to explain why for DiR NCs a significantly lower DCP is observed in magnetic fields compared to spherical NCs. The observed dependence on the aspect ratio highlights the role of the dielectric enhancement effect, which favors emission polarized parallel to the c-axis in DiRs. The reason for the additional increase of the ratio $\Gamma_{20}/\Gamma_{21}^0$ with increase of the DiR aspect ratio can be explained by the increase of the shape anisotropy in the plane perpendicular to the rod axis or by the increase of the
Coulomb interaction anisotropy. Admittedly, the suggested model has a few free parameters and the theoretical prediction of the linear and circular polarization properties for DiR have to be considered with care. Additional dedicated measurements at low temperatures are necessary to elucidate the relevance and the magnitude of each parameter and their dependences on NC size and shape. Especially the determination of the linear polarization properties of individual DiRs at low temperatures and in magnetic fields would be highly enlightening in this regard. Nevertheless, we believe, that these findings may lead to a deeper understanding of the processes related to the exciton fine structure and the resulting polarization properties of the emitted light. Finally, we would like to point out, that also for spherical core/shell NCs saturation DCP values considerably lower than the theoretically expected $-0.75$ have been observed in high magnetic field experiments and these low DCPs could not be explained at that time.\cite{Furis2005,Wijnen2008} However, they could be explained in the framework of our theory, if one considers, that also in spherical core/shell NCs the coupling to the $F=0^U$ state is relevant. This would intrinsically lower the saturation value of the DCP in the ensemble as discussed in connection with Fig.~\ref{figure13}. Furthermore, real NCs are never perfectly spherical, which makes it necessary to consider also the dielectric enhancement effect for these structures.

\section{Conclusions}
\label{outlook}

In summary, we performed a comprehensive study of the PL polarization and exciton spin dynamics in CdSe/CdS core/shell NCs in external magnetic field. We demonstrated the possibility of probing the orientation of NCs with anisotropic shell shape with respect to the magnetic field axis by measuring the PL DCP induced by the field, and highlighted possible applications of this technique for testing nanorod based arrays.

Regarding exciton spin dynamics we found that the spin relaxation of dark excitons is strongly related to the surface of CdSe in bare NCs or the interface between CdSe and CdS in core/shell systems. This was evidenced by a large decrease of the spin relaxation rate after adding a CdS shell. The importance of the interface in core/shell systems was highlighted by a strong core size dependence of the spin relaxation rate, which decreases for increasing core volume. In contrast, the overall NC surface seems to play only a minor role in this regard, as concluded from the weak dependence of the spin relaxation rate on shell thickness. The temperature and magnetic field dependences of the spin relaxation rate emphasize the importance of confined acoustic phonons in the relaxation and recombination of dark excitons. In this respect, DiR NCs show a similar behaviour as spherical core/shell systems. We believe, that these findings may stimulate further discussion about the role of the core/shell interface for the optical properties, in particular the dark exciton recombination and spin relaxation in NCs. With respect to applications our results on spin dynamics may help to figure out the relevant parameters for spin-preserving environments in spintronics devices, e.g. for spin storage. In respect thereof we can state, that adding a shell for surface passivation and the use of large core NCs as well as thick shells favor long spin lifetimes.

Finally, we extensively studied the circular PL polarization of DiRs in magnetic field. Despite a remarkably low equilibrium DCP for randomly oriented DiRs compared to spherical core/shell NCs our results point towards the dependence of the saturated DCP on the NC aspect ratio. We elaborated a theory for analyzing and predicting the circular and linear polarization properties of DiR emission. It includes the dielectric enhancement effect as well as dark exciton recombination via coupling to bright exciton states with radiative transitions polarized parallel or perpendicular to the hexagonal c-axis. This model can explain the strongly varying (almost saturated) DCPs typically observed even for spherical NCs in high magnetic fields and may lead to a better understanding of the exciton fine structure and its relevance for the PL polarization properties of NCs especially at low temperatures.

\section{Acknowledgment}
The authors thank Al.L. Efros for discussions.  This work was supported by the Government of Russia (Project No.14.Z50.31.0021, leading scientist M. Bayer), The Deutsche Forschungsgemeinschaft (Sonderforschungsbereich TRR160), the EU Seventh Framework Programme (Grant No. 214954, HERODOT) and BelSPo (IAP7.35 - photonics$@$be). A.V. Rodina acknowledges support of the Russian Foundation for Basic Research (Grant No. 13-02-00888).


%


\end{document}